\documentclass[aps,pre,twocolumn,groupedaddress,showpacs]{revtex4}
\usepackage{graphicx}
\usepackage{amssymb}

\begin{document}


\title{Speckle-visibility spectroscopy: A tool to study time-varying dynamics}



\author{R. Bandyopadhyay}
\altaffiliation[Permanent address: ]{Liquid Crystals Laboratory,
Raman Research Institute, Bangalore 560080, INDIA}
\email{<ranjini@rri.res.in>.}
\author{A.S. Gittings}
\author{S.S. Suh}
\author{P.K. Dixon}
\altaffiliation[Permanent address: ]{Department of Physics and
Astronomy, California State University, San Bernardino, CA 92407,
USA} \email{<pdixon@csusb.edu>.}
\author{D.J. Durian}
\altaffiliation[Permanent address: ]{Department of Physics and
Astronomy, University of Pennsylvania, Philadelphia, PA 19104,
USA} \email{<djdurian@physics.upenn.edu>.} \affiliation{Department
of Physics and Astronomy, University of California, Los Angeles,
CA 90095, USA.}


\date{\today}

\begin{abstract}
We describe a multispeckle dynamic light scattering technique
capable of resolving the motion of scattering sites in cases that
this motion changes systematically with time. The method is based
on the visibility of the speckle pattern formed by the scattered
light as detected by a single exposure of a digital camera.
Whereas previous multispeckle methods rely on correlations between
images, here the connection with scattering site dynamics is made
more simply in terms of the variance of intensity among the pixels
of the camera for the specified exposure duration.  The essence is
that the speckle pattern is more visible, i.e.\ the variance of
detected intensity levels is greater, when the dynamics of the
scattering site motion is slow compared to the exposure time of
the camera.  The theory for analyzing the moments of the spatial
intensity distribution in terms of the electric field
autocorrelation is presented. It is demonstrated for two
well-understood samples, a colloidal suspension of Brownian
particles and a coarsening foam, where the dynamics can be treated
as stationary. However, the method is particularly appropriate for
samples in which the dynamics vary with time, either slowly or
rapidly, limited only by the exposure time fidelity of the camera.
Potential applications range from soft-glassy materials, to
granular avalanches, to flowmetry of living tissue.
\end{abstract}



\maketitle



\section{Introduction}

Dynamic light scattering (DLS) is a powerful tool for probing
motion within samples of physical, chemical, biological, and
medical interest \cite{CumminsPike74, BernePecora, Chu, Brown,
SheperdOberg, asakura91, BriersPM01}. The physical basis is that
the frequency spectrum of the scattered light is Doppler broadened
according to the velocities of all the scattering sites. The shape
of the spectrum reveals the nature of the motion, for example
whether it is ballistic or diffusive; the characteristic width of
the spectrum reveals the rate of the motion, for example the
root-mean-squared speed or the diffusion coefficient. If the
sample is nearly transparent, so that incident photons scatter at
most once, then the spectrum can be resolved vs scattering angle
in order to probe collective motion at different length scales.
This is the single-scattering regime. By contrast if the sample is
opaque, so that incident photons scatter off many sites before
exiting the sample, then any wavevector-dependence is lost. The
art of DLS in this regime is known as diffusing-wave spectroscopy
\cite{MaretWolf87, PineDWS88, DWSrev93, DWSrev97}.

The most straightforward approach to DLS is to measure the
frequency spectrum directly, for example using a Fabry-Perot
interferometer with a very narrow band pass.  However, it is also
common practice to deduce the spectrum by an interference
technique, in which the scattered light is collected over an area
comparable to one speckle spot (spatial-coherence length).  The
motion of the scattering sites causes corresponding changes in the
speckle pattern, and hence large fluctuations in the detected
intensity. These fluctuations are quantified by the temporal
intensity autocorrelation function, which is simply related to the
frequency spectrum under certain conditions (below). This is known
as intensity- or photon-correlation spectroscopy (PCS). One
advantage of PCS is that digital correlators are commercially
available that can compute the intensity autocorrelation over many
decades in delay time, for example 10~ns to 100~s. One
disadvantage of this approach is that the temporal fidelity is
limited by the necessity of sampling over many correlation times
to build up statistical weight. This makes them a poor choice for
studying systems with dynamics changing on time scales of seconds
or faster. Interferometers are useful for large frequency shifts,
but do not sport an equally impressive dynamic range. Given the
breadth of applications of DLS, it is perhaps not surprising that
the essential equivalence of information available from
interferometric and correlation-based approaches to DLS is not
universally recognized \cite{BriersJOSA96}.

In order to ensure simple connection between the intensity
autocorrelation and the frequency spectrum of the scattered light,
several conditions must be met: (a) There must be many,
uncorrelated scattering sites or regions; (b) the extent of the
motion must be sufficiently great as to fully randomize the
speckle pattern; and (c) the scattering site dynamics must not
vary over the time scale of the measurement. The first criterion
holds if the sample and scattering volume are sufficiently large;
this does not represent a fundamental restriction. The second
criterion holds if the sample is fluid or if the scattering sites
are bound only loosely to a fixed average location.  The third
criterion holds if the sample is in thermal equilibrium, or if the
sample is in a stationary state in which both the external energy
input and the microscopic dynamical response do not fluctuate.
Thus, the conditions (a-c) for conventional single-detector PCS to
apply are not overly restrictive.  It is possible to verify
whether not these conditions hold through measurement of
higher-order temporal intensity correlations \cite{PierreJOSA99},
which can be processed from the raw intensity vs time data stream
simultaneously with the second-order intensity autocorrelation.

There are many systems where some of the above conditions do not
hold and conventional single-detector PCS does not apply. The
kinetics of phase separation, gelation, and aggregation are
examples of long-standing interest, in which the dynamics
progressively change with time \cite{PoonCOCIS98}. These processes
can be treated as stationary only if the evolution is slow
compared to the time scale over which the intensity
autocorrelation decays. The broad class of soft-glassy materials
comprise another example where the dynamics change with time
\cite{LucaRev05}. Furthermore, just as for the gelation problem,
the scattering sites can become more tightly bound with age, so
that after a certain point the speckle pattern no longer fully
randomizes.  And lastly, dynamics in granular materials usually
cannot be studied with traditional PCS, for example because the
input of energy is vibratory or because the response is
intermittent avalanche-like flows \cite{JNBrev96}.

Multispeckle dynamic light scattering techniques have been
introduced as a useful remedy in such situations where traditional
single-detector PCS methods do not apply \cite{wongRSI, Kirsch96}.
The approach is to compute the temporal autocorrelation function
for each pixel of a digital camera, and then to average together
the results. Since there are many pixels, and hence many speckles,
it is no longer a requirement that motion within the sample cause
the speckle pattern to fully randomize. And since the large number
of pixels can significantly reduce the time needed to acquire good
signal-to-noise, it is easier to study evolving dynamics.  However
it still remains a challenge to implement multispeckle DLS. A
prohibitive difficulty is that commercial multispeckle devices do
not exist.  A limiting difficulty is either that vast quantities
of data must be stored for post-processing or that real-time
processing must be made sufficiently fast.  Further difficulties
arise from the use of charge coupled devices as light sensing
elements.  Hardware and software advances continue to be reported
in the technical literature \cite{lucaRSI99, mochrieRSI00,
pineRSI02, xu02, lucaTRC, GrubelRSI03, PuseyRSI04, mochrieRSI04}.

In this paper we supply full details and demonstration of a
multispeckle dynamic light scattering technique we dub
speckle-variance spectroscopy \cite{PaulBAPS} or
speckle-visibility spectroscopy (SVS) \cite{SVS}. Our approach is
to characterize motion within a sample in terms of the visibility
of the speckle pattern formed with scattered light for a single
exposure of a digital CCD or CMOS camera.  We begin by introducing
appropriate notation and the experimental apparatus in the context
of the more usual multispeckle DLS.  Then we describe the
theoretical underpinnings of SVS, and give examples for common
types of scattering site dynamics.  Our theory contradicts a
widely-cited prediction obtained in the context of laser-speckle
photography \cite{BriersOC81}.  Next, crucially, we demonstrate
the validity of our theory by experiments on well-known samples.
Finally we discuss experimental considerations for successful
implementation of dynamic light scattering with a digital camera.

\section{Photon-Correlation Spectroscopies}

We begin with prerequisite theoretical and experimental background
materials necessary for the next sections on SVS.

\subsection{Theory}

In all DLS experiments, light from a coherent source enters a
sample. Some portion scatters, with individual photons
experiencing different trajectories, and some fraction of the
scattered light reaches a photodetector.  Ignoring constant
factors, the detector reports a signal proportional to the light
intensity, $I(t)=E(t)E^{*}(t)$, where the electric field $E(t)$ is
a superposition of many fields representing many photon
trajectories.  The acquired intensity can be an analog signal, or
it can be a bit-stream with each pulse representing a different
detected photon.  Ultimately, the quantity of interest is either
the power spectrum $|E(\omega)|^2$ or its Fourier transform: the
temporal electric field autocorrelation function.  We denote the
absolute normalized temporal electric field autocorrelation as
\begin{equation}
    g_1(\tau) \equiv |\langle E(t)E^*(t+\tau)\rangle|
        / \langle E(t)E^*(t)\rangle,
\label{g1}
\end{equation}
where $\tau$ is the delay time.  In traditional PCS the average
$\langle\cdots\rangle$ is taken over a range of times $t_{\rm
start}<t<t_{\rm stop}$. By definition, $g_1(\tau)$ decays from one
to zero as $\tau$ ranges from zero to infinity. The characteristic
time scale for the decay is the reciprocal of the characteristic
width of the power spectrum. If the power spectrum is symmetric
and centered around $\omega_\circ$, then the normalized (but not
absolute) electric field autocorrelation function is
$g_1(\tau)e^{i\omega_\circ\tau}$.  For example, a Lorentzian power
spectrum $|E(\omega)|^2 \propto 1/[Dq^2+(\omega-\omega_\circ)^2]$
and an exponential field autocorrelation function
$g_1(\tau)=\exp(-Dq^2\tau)$ correspond to light of incident
frequency $\omega_\circ$ scattered by wavevector $q$ from
diffusing particles; the diffusion coefficient $D$ could be
extracted from measurement of either the power spectrum or the
field autocorrelation.

In single-detector PCS the electric field autocorrelation function
is deduced from measurement of the normalized intensity
autocorrelation function,
\begin{equation}
    g_2(\tau) \equiv \langle I(t)I(t+\tau)\rangle/\langle I\rangle^2.
\label{g2}
\end{equation}
This is straightforward only if the three conditions discussed in
the Introduction are all met.  If (a) the electric field is the
superposition of many independent scattered fields and if (b) the
field autocorrelation decays to {\it zero} over a time scale much
shorter than the duration of the measurement, then the Central
Limit Theorem implies that $E(t)$ is a Gaussian-distributed
complex variable with zero mean.  Intuitively, the total field
$E(t)=\sum E_i(t)$ at some instant of time, $t$, may be evaluated
graphically by phasor addition. If there are enough independent
scattering regions, then each term in the sum constitutes one step
in a random walk in the complex plane.  Many such random walks
will be sampled, and hence the distribution of values of $E(t)$
over the course of the measurement will be Gaussian, if
$g_1(\tau)$ fully decays to zero over a shorter time scale than
the measurement duration.  If the field distribution is Gaussian,
then temporal correlations of the form $\langle E(t)E^*(t)
E(t+\tau_1)E^*(t+\tau_1) E(t+\tau_2)E^*(t+\tau_2)\cdots\rangle$
can be expressed as a sum of products of field autocorrelations.
For example, the normalized intensity autocorrelation is a
four-order field correlation that reduces to
\begin{equation}
    g_2(\tau)=1+\beta[g_1(\tau)]^2,
\label{siegert}
\end{equation}
where $\beta\le1$ is a number determined by the ratio of detector
size to speckle spot size. This is widely known as the Siegert
relation. A detailed derivation of the Siegert relation, the value
of $\beta$, and analogous results for third- and fourth-order
temporal intensity correlations, are given in
Ref.~\cite{PierreJOSA99}.  To briefly summarize, the method of PCS
is to measure $g_2(\tau)$ and to extract $g_1(\tau)$ using
Eq.~(\ref{siegert}).  Subsequent connection is then to be made
between $g_1(\tau)$ and scattering site dynamics, depending on
details of the illumination and detection geometry and on the
optical properties of the sample.

In multispeckle PCS methods the intensity autocorrelation is
measured at each pixel of a digital camera and the results are
averaged together \cite{wongRSI, Kirsch96,lucaRSI99, mochrieRSI00,
pineRSI02, xu02, lucaTRC, GrubelRSI03, mochrieRSI04}.  By virtue
of the large number of pixels, the combined statistics of all the
detected fields is now Gaussian even if the field autocorrelation
never decays to zero.  In effect, statistics are sampled by an
ensemble average over many speckles rather than by a time-average
for a single speckle. Thus the Siegert relation,
Eq.~(\ref{siegert}), may be invoked even more generally for
multispeckle PCS than for single-detector PCS.

As an aside, the violation of the Siegert relation in
single-detector PCS due to non-randomization of the detected
electric field is sometimes said to be due to non-ergodicity of
the sample.  This is a misnomer and can lead to confusion.  The
ergodicity of dynamics within the sample, and the ergodicity of
the field statistics for the detected light, are distinct issues
that may or not be related.

\subsection{Experiment}

We now apply the above multispeckle PCS technique to a suspension
of diffusing Brownian particles.  This serves as a starting point
from which to demonstrate SVS, since all measurement and sample
hardware carry over without change.

\begin{figure} 
\includegraphics[width=\columnwidth]{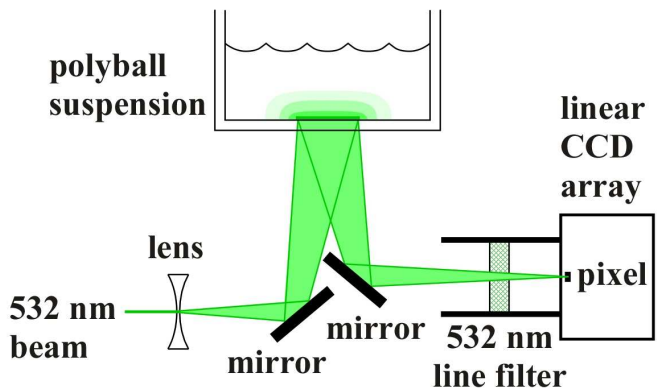}
\caption{Experimental apparatus for performing both multispeckle
dynamic light scattering and speckle-visibility spectroscopy on an
opaque colloidal suspension.  The orientation of the array of
pixels is horizontal, perpendicular to the plane of the drawing.}
\label{setup}
\end{figure}

The sample consists of 653~nm diameter polystyrene spheres [Duke
Scientific] suspended in water at a volume fraction of ten
percent. It is poured into a glass beaker, diameter 6~cm, to a
depth of 2.4~cm, then sealed.  Light from a Coherent Verdi V5
NdYVO$_4$ Laser, wavelength $\lambda=532$~nm, is expanded and
directed almost normal to the bottom of the sample beaker with a
Gaussian spot size of $a=1.25$~cm. The outpower of the laser is
held fixed, and the illumination intensity is reduced as needed by
use of neutral density filters. See the schematic diagram in
Fig.~\ref{setup}. According to Mie scattering theory for dilute
independent spheres \cite{vandeHulst}, the scattering length
specifying the exponential attenuation of a beam is $l_s=24~\mu$m,
and the average cosine of the scattering angle is $g=0.90$.
Therefore, about ten scattering events are required to randomize
the photon propagation direction, and the transport mean free path
is $l^*=l_s/(1-g)\approx 240~\mu$m. Thus our sample has an opaque
white appearance, and we operate in the multiple scattering regime
known as diffusing-wave spectroscopy.

In order to perform multispeckle DWS, a portion of the {\it
backscattered} light leaving the bottom of the sample is reflected
by mirror into a Basler-160 digital line scan CCD camera.  This
device has 1024 pixels, each $10~\mu$m $\times$ $10~\mu$m and 8
bits deep, and can capture images at a maximum rate of 58~kHz.
Except for the mirror and a 532~nm line filter, there are no other
optics. The sample-to-camera distance is adjusted to about
$d=30$~cm.  This gives a speckle size of $s \approx d \lambda /
a=13~\mu$m, and a ratio of pixel to speckle areas of $A_{\rm
pixel}/A_{\rm speckle}\approx 0.6$. The camera is interfaced to a
PC equipped with a National Instruments PCI-1422 card, and is
programmed using LabVIEW.

\begin{figure} 
\includegraphics[width=\columnwidth]{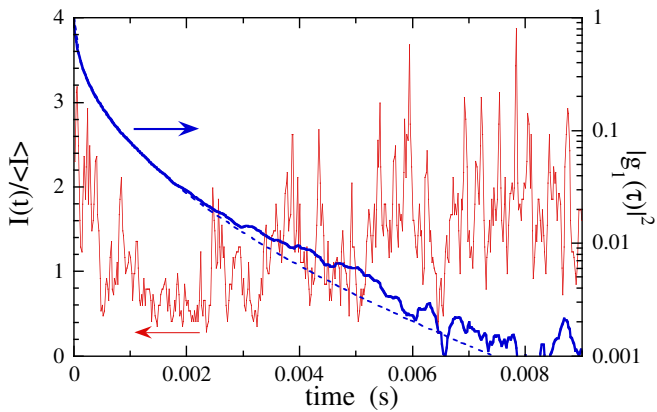}
\caption{An example of intensity vs time in one pixel (left) and
the normalized intensity autocorrelation function
$|g_1(\tau)|^2=[\langle I(0)I(\tau)\rangle/\langle
I\rangle^2-1]/\beta$ (right), for light backscattered from an
opaque colloidal suspension.  The intensity autocorrelation is
directly computed from the full time-trace, which consists of
$10^5$ points in 20~$\mu$s time increments for each of the 1024
pixels in the linescan CCD camera. The normalization factor
$\beta=0.34$ was obtained by extrapolating the unnormalized
intensity autocorrelation to zero delay time.  The suspension
consists of 653~nm diameter polystyrene spheres in water, at a
volume fraction of 10 percent. The theory of diffusing-wave
spectroscopy predicts $g_1(\tau)\sim\exp[-\gamma\sqrt{6t/t_o}]$
where $t_o=1/Dk^2=6.1$~ms.  The fit to this form (dashed curve)
gives $\gamma=1.3$. } \label{Ig1}
\end{figure}

As a benchmark reference to compare with our SVS technique, our
operating procedure is to record the intensity levels in all
1024 pixels for a total of 2~s in increments of $20~\mu$s; the
entire data set thus consists of 102.4 million 8-bit values. The
laser intensity is adjusted so that the average gray-scale value
is 40. When the laser is blocked, the signal drops to a ``dark
count'' grayscale value of 3.5.  The first step in the analysis is to
subtract the dark count and divide by the average remaining
signal, thus giving the normalized intensity time trace
$I(t)/\langle I\rangle$ for each pixel.  A portion of such a trace
for one pixel is displayed in Fig.~\ref{Ig1}.  As the colloidal
particles diffuse, the intensity level at a pixel indeed fluctuate
strongly; here, it is seen to vary between about 0.4 and 4 times
the average.

The observed intensity fluctuations in Fig.~\ref{Ig1} display
structures lasting over a range of time scales.  As is done in
multispeckle PCS, this behavior may be quantified by the
normalized intensity autocorrelation, $g_2(\tau)$ defined by
Eq.~(\ref{g2}), which we compute directly for each pixel and then
average together. According to the Siegert relation,
Eq.~(\ref{siegert}), the zero-time intercept is $g_2(0)=1+\beta$.
Extrapolating $g_2(\tau)$ data to $\tau=0$ gives $\beta=0.34$,
which is consistent with the ratio of pixel to speckle areas.
Invoking the Siegert relation, we deduce the normalized field
autocorrelation $|g_1(\tau)|$, and plot its square in
Fig.~\ref{Ig1}.  Finding the value of $\beta$ is often referred to
as the issue of normalization.  Note that the square of
$g_1(\tau)$ is simply the intensity autocorrelation displayed
dimensionlessly on a scale ranging from 1 to 0.  Thus, the time
scales of structure in the intensity time trace can be compared
directly with features in the decay of $|g_1(\tau)|^2$. Indeed the
decay is very fast initially, reflecting the fast fluctuations in
$I(t)$. The later-time decay is slower, reflecting the
longer-lived fluctuations evident in Fig.~\ref{Ig1} as drift in a
local average of the intensity.

The above measurement of $g_1(\tau)$ may be compared with the
predictions of DWS.  In the backscattering geometry, with
equivalent plane-wave in / plane-wave out illumination and
detection, the theory of DWS \cite{DWSrev93} predicts
$g_1(\tau)\approx\exp(-\gamma\sqrt{6\tau/\tau_\circ})$ where
typically $1.5<\gamma<2.5$ and where $\tau_\circ \equiv 1/(Dk^2)$
is the characteristic time for a particle to diffuse a distance
$1/k=\lambda/(2\pi n)$ where n is the index of refraction. For our
sample, the predicted decay time is $\tau_\circ=6.1$~ms. The
stretched-exponential form of $g_1(\tau)$ reflects the broad
length distribution of possible photon paths that contribute to
the signal.  It also reflects a subtle breakdown of diffusion and
continuum approximations for short path lengths
\cite{Middleton91,DWSacc95}.  The value of $\gamma$, but not the
stretched-exponential form, is particularly sensitive to the
treatment of short paths and can thus be affected by the
polarization states, boundary reflectivities, and propagation
directions for the incoming and outgoing photons
\cite{Fred89,MoinJOSA97,PierrePRE98}.  Taking the value
$\gamma=1.28$, somewhat lower than expectation, we obtain an
excellent fit to the data as shown by the dashed curve in
Fig.~\ref{Ig1}. Thus we pronounce our sample, apparatus, dataset,
and multispeckle analysis methods as sound. For demonstration of
SVS in a later section, the value of $\gamma$ will not be
important; we only need a sample with known $g_1(\tau)$.

\section{Speckle-visibility spectroscopy}

\begin{figure} 
\includegraphics[width=\columnwidth]{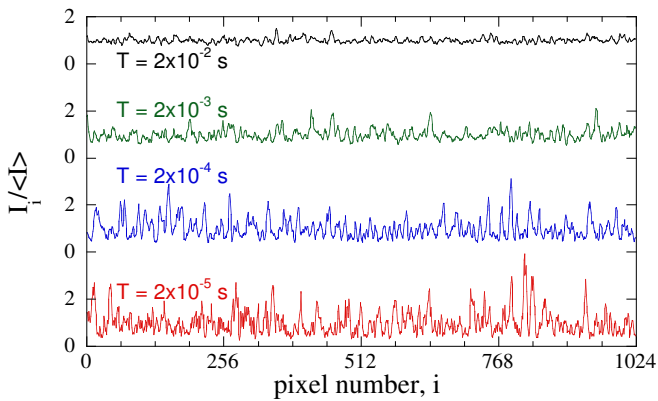}
\caption{Intensity vs pixel number, ie the profile of the speckle
pattern in the plane of the CCD camera, for the same colloidal
suspension and optical configuration as in Figure 1. The exposure
durations $T$ differ by successive factors of ten, as labelled.
Since the speckles change with time, as shown in
Fig.~\protect{\ref{Ig1}}, their visibility is smaller for longer
exposures.  This is the essence of SVS.}\label{IT}
\end{figure}

In this section we develop the theory of speckle-visibility
spectroscopy (SVS).  The underlying principle of SVS is
illustrated in Fig.~\ref{IT}, which displays intensity vs pixel
number for four different exposure times, $T$, of the camera.  The
shortest exposure in Fig.~\ref{IT}, $T=2\times10^{-5}$~s, is
shorter than the decay time of $g_1(\tau)$; therefore, the speckle
appears static and large intensity differences are registered from
pixel to pixel. For longer and longer exposures, the visibility of
this speckle pattern progressively fades. This is because the
intensity at each individual pixel fluctuates during the exposure
and is averaged over the exposure time. In the limit of a very
long exposure time in comparison to the decay time, each pixel
approaches the same mean intensity value and there is no variation
among the pixels.  Indeed, the longest exposure in Fig.~\ref{IT},
$T=2\times10^{-2}$~s, is longer than the decay time of
$g_1(\tau)$; here, many speckles are sampled at each pixel over
the duration of the exposure, and each pixel registers a value
close to the average.  The very essence of SVS is, thus, to
quantify the visibility of the speckle pattern in terms of moments
of the distribution of intensities registered by all the pixels
for a given exposure duration, and to relate this to the absolute
normalized electric field autocorrelation function $g_1(\tau)$.  A
subsequent connection with scattering site dynamics can then be
made as per usual DLS practice in either single- or multiple
scattering limits.

Before carrying out the theoretical aspects of this program, we
note that our method is not without precedent.  Perhaps the first
is a calculation \cite{jakeman68a} and experimental verification
\cite{jakeman68b} of the distribution for the photocurrent as
measured by one detector as a function of integration time.
Another precedent is ``laser-speckle photography''
\cite{BriersOC81}, in which the blurring of speckle in a
laser-illuminated scene is taken as a signature of motion
\cite{asakura91,BriersPM01}.  The latter is now being applied to
cerebral blood flow, in particular \cite{DunnJCBFM01,
DurduranJCBFM04, WeberEJN04, YuanAO05}.  One aspect of our
contribution here is to simplify and generalize the work of
Refs.~\cite{jakeman68a}, and to correct a mistake in the
widely-cited work of Ref.~\cite{BriersOC81}.

\subsection{Variance}

The variance of intensity across the pixels is a simple way to
quantify the visibility of the speckle pattern formed at the imaging
array. For a given exposure, each pixel reports a signal that is
proportional to the total number of photons it receives. Thus the
signal at pixel $i$ is proportional to the time-average of the
intensity trace $I_i(t)$:
\begin{equation}
    S_{i,T}=\int_{0}^{T}I_i(t'){\rm d}t'/T,
\label{Si}
\end{equation}
where $t=0$ defines the beginning of the exposure and $T$ is the
duration of the exposure. The data returned by the camera, for a
single exposure, consists of the set $\{S_{i,T}\}$ where the index
$i$ ranges from 1 to the total number $N$ of pixels.  All
quantities of interest are to be computed from the $N$ members of
this set. For example the $n^{th}$-moment of the distribution of
pixel signals is
\begin{equation}
    \langle I^n\rangle_T = \sum_{i=1}^{N}(S_{i,T})^n/N,
\label{Sn}
\end{equation}
where the subscript $T$ is a reminder that the result depends on
the exposure duration. Note that these moments represent an {\it
ensemble} average over pixels for a fixed time interval.

To compute the variance we focus on the first two moments of the
signal distribution.  The first moment is simply the average
intensity, $\langle I\rangle=\sum_{i=1}^{N}S_{i,T}/N$, which is
independent of the exposure duration.  The second moment is the
average over pixels of the quantity
\begin{equation}
    (S_{i,T})^2 = \int_0^T\int_0^T I_i(t')I_i(t''){\rm d}t'{\rm
    d}t'' / T^2.
\label{S2}
\end{equation}
Since this is an {\it ensemble} average, the Siegert relation
Eq.~(\ref{siegert}) may be invoked: $\langle
I_i(t')I_i(t'')\rangle = \langle I\rangle^2
\{1+\beta[g_1(t'-t'')]^2\}$, giving an intermediate result for the
second moment as
\begin{equation}
    \langle I^2 \rangle_T = \langle I \rangle^2
     \int_0^T\int_0^T  \{1+\beta [g_1(t'-t'')]^2\}{\rm d}t'{\rm d}t'' /
     T^2.
\label{I2}
\end{equation}
The first term in the integral is one; the second term can be
reduced to a single integral by recognizing that $g_1(t)$ is
usually an even function.  We now define a normalized variance,
and finish the calculation:
\begin{eqnarray}
    V_2(T) &\equiv& {1\over\beta}
            \left[\langle I^2\rangle_T/\langle I\rangle^2 -1 \right],\cr
    &=&  \int_0^T\int_0^T [g_1(t'-t'')]^2 {\rm d}t'{\rm d}t'' / T^2,\cr
    &=& \int_0^T 2(1-t/T)[g_1(t)]^2 {\rm d}t/T.
\label{V2}
\end{eqnarray}
This is the fundamental equation of SVS.  The top line is a
definition; it quantifies speckle visibility on a scale of $0-1$
in terms of the first two moments of the distribution of pixel
signal data, $\{S_{i,T}\}$, returned for a given exposure of
duration $T$.  The middle line is an intermediate step that holds
even if $g_1(t)$ is not even. The bottom line is where contact
usually is to be made between measurement and the underlying
normalized electric field autocorrelation. Evidently the variance
is a weighted-average of $[g_1(t)]^2$ over the exposure interval
$0<t<T$, with heavier weighting for shorter $t$.  This weighting
reflects the distribution of possible time differences within an
exposure.

\subsection{Higher-Order Moments}

The distribution of pixel signals is typically skewed toward
higher values, as seen for example in Fig.~\ref{IT}; therefore, it
is not Gaussian and cannot be fully specified by just the value of
the variance.  Hence we repeat the calculation leading to the
fundamental equation of SVS, Eq.~(\ref{V2}), but now for
higher-order moments.  The results can be useful for diagnosing
problems with the experimental apparatus, for deducing the
normalization constant $\beta$, and for better testing trial forms
of $g_1(t)$ for unknown samples.

We define reduced moments of the pixel signal distribution as
\begin{equation}
    v_n(T) \equiv \langle I^n \rangle_T / \langle I\rangle^n-1.
\label{vndef}
\end{equation}
These are larger for shorter exposures, and vanish for long
exposures in the limit that the speckle is no longer visible.
While dimensionless, these moments are not normalized in the sense
that their values depend on the number of speckles per pixel
through $\beta$.  Hence we use a lower-case ``$v_n(T)$'' for {\it
reduced} moments defined in Eq.~(\ref{vndef}), to contrast with
the upper-case ``$V_2(T)$'' for the {\it normalized} second moment
defined in Eq.~(\ref{V2}). Now the task is to compute an ensemble
average over pixels $i$ for quantities of form
\begin{widetext}
\begin{equation}
  \left\langle \int_0^T{{\rm d}t_1\over T} \int_0^T{{\rm d}t_2\over T}
    \cdots \int_0^T{{\rm d}t_n\over T}
    I_i(t_1)I_i(t_2)\cdots I_i(t_n) \right\rangle_i.
\label{In}
\end{equation}
Invoking third and fourth order Siegert relations
\cite{PierreJOSA99}, and assuming that $g_1(t)$ is even, we arrive
at
\begin{eqnarray}
    v_2(T) &=& 2\beta\int_0^T{{\rm d}t\over T}\left(1-t/T\right)[g_1(t)]^2,\cr
    v_3(T) &=& 3v_2(T)+12\beta^2
        \int_0^T{{\rm d}t_1\over T}
        \int_{t_1}^T{{\rm d}t_2\over T}
        \left(1-t_2/T\right)g_1(t_1)g_1(t_2)g_1(t_2-t_1),\cr
    v_4(T) &=& 6v_2(T)+O(\beta^2)
\label{v234}
\end{eqnarray}
\end{widetext}
The reduced second moment is recognized as $v_2(T)=\beta V_2(T)$.
The third and fourth reduced moments, by contrast, contain terms
that are not proportional to $\beta$; therefore, their dependence
on the number of speckles per pixel cannot be normalized away by a
simple division.

Note that for very short exposure times, the reduced moments
approach
\begin{equation}
    v_n(0)=[1+\beta][1+2\beta]\cdots[1+(n-1)\beta]-1,
\label{vn0}
\end{equation}
which is a well-known result describing the moments of static
speckle patterns \cite{PierreJOSA99, Goodman84}.  For long
exposure times, the reduced moments vanish as
\begin{eqnarray}
    v_2(T) &\rightarrow& 2\beta
        \int_0^\infty[g_1(t)]^2{\rm d}t/T,\cr
    v_3(T) &\rightarrow& 3v_2(T),\cr
    v_4(T) &\rightarrow& 6v_2(T).
\label{vnlong}
\end{eqnarray}
No matter what the form of $g_1(t)$, the final decay of the
moments goes as $v_n(T)\sim1/T$.  This is a consequence of the
heavy weighting of $g_1(t)$ near $t=0$.

\subsection{Examples and Precedents}

\begin{figure} 
\includegraphics[width=\columnwidth]{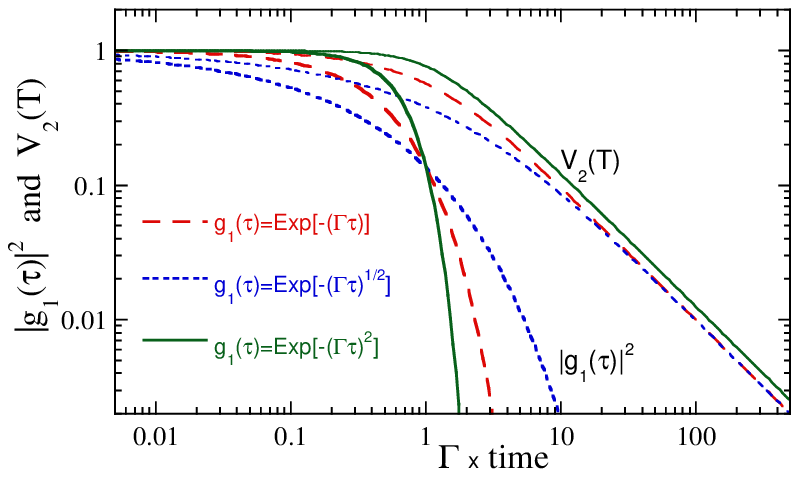}
\caption{Comparison of $[g_1(\tau)]^2$ and $V_2(T)$, for three
types of scattering site dynamics, as labelled. According to
prediction, Eq.~(\protect{\ref{V2}}), the latter is a weighted
average of the former.} \label{g2v2theory}
\end{figure}

Here we consider the intensity moments predicted by the above
formalism for several forms of $g_1(\tau)$ of experimental
interest. Normalized variance predictions for five special cases
are collected in Table~\ref{V2table}.  The first three of these
cases are plotted vs exposure time in Fig.~\ref{g2v2theory}.  Note
that the decay of $V_2(T)$ vs $T$ is slower than that of
$g_1(\tau)$ vs $\tau$; furthermore, the long-time decay is always
$V_2(T)\sim1/T$, as explained by Eq.~(\ref{vnlong}). This feature
allows the characteristic time scale for the decay of $g_1(\tau)$,
which is equivalent to the characteristic broadening of the power
spectrum, to be extracted even when it is a decade or more faster
than the bandwidth of the camera.

\begin{table}
\begin{ruledtabular}
\begin{tabular}{ll}
$g_1(x=\Gamma \tau)$ & $V_2(x=\Gamma T)$ \\ \colrule
$\exp(-x)$ & $[ e^{-2x}-1+2x ]/(2x^2)$ \\
$\exp(-\sqrt{x})$ &
    $[ (3+6\sqrt{x}+4x)e^{-2\sqrt{x}}-3+2x]/(2x^2)$ \\
$\exp(-x^2)$ & $[e^{-2x^2} - 1 + \sqrt{2\pi}x{\rm erf}(\sqrt{2}x) ]/(2x^2)$ \\
$\sqrt{6x}/\sinh{\sqrt{6x}}$ & $\zeta(3)/x\approx1.202/x$, $x\gg1$ \\
$\sqrt{6x^2}/\sinh{\sqrt{6x^2}}$ & $\sqrt{\pi^4/54}/x\approx1.343/x$, $x\gg1$ \\
\end{tabular}
\end{ruledtabular}
\caption{Normalized variance predictions, computed from
Eq.~(\protect{\ref{V2}}), for various forms of the normalized
electric field autocorrelation function, $g_1(\tau)$. The first
example, $g_1(\tau)=\exp(-\Gamma\tau)$, corresponds to
single-scattering from sites with diffusive dynamics and to DWS in
backscattering for a sample with ballistic dynamics.  The second
example, $g_1(\tau)=\exp(-\sqrt{\Gamma\tau})$, corresponds to DWS
in backscattering for a sample with diffusive dynamics.  The third
example, $g_1(\tau)=\exp[-(\Gamma\tau)^2]$, corresponds to
single-scattering for a sample with ballistic dynamics.  These
three cases are plotted vs exposure time in
Fig.~\protect{\ref{g2v2theory}}. The fourth and fifth examples
correspond to DWS in transmission for samples with diffuse and
ballistic dynamics, respectively. These forms are too intractable
to compute for general exposure times.} \label{V2table}
\end{table} 

The only case for which we have analytically computed both second
and third moments of the pixel signal distribution is for a
Lorentzian spectrum or, equivalently, for an exponential field
autocorrelation $g_1(\tau)=\exp(-\Gamma\tau)$. This corresponds to
single-scattering from a sample with diffusive dynamics and to DWS
in backscattering from a sample with random ballistic dynamics.
For the former, the linewidth or decay rate is $\Gamma=Dq^2$ where
$D$ is the diffusion coefficient and $q$ is the magnitude of the
scattering vector; for the latter, the decay rate is
$\Gamma\approx 4\pi \delta v/\lambda$ where $\delta v$ is the
root-mean squared average random speed and $\lambda$ is the
wavelength of light in the medium.  For this example, the reduced
second and third moments are
\begin{eqnarray}
    v_2(T) &=& \beta{e^{-2x}-1+2x \over 2x^2 },\cr
    v_3(T) &=& 3v_2(T)+6\beta^2{(1+x)e^{-2x}-1+x \over 2x^3},
\label{v23exp}
\end{eqnarray}
where $x=\Gamma T$ is the product of decay rate and exposure time,
as per the notation in Table~\ref{V2table}.

The normalized variance for the special case of a Lorentzian
spectrum, given in Eq.~(\ref{v23exp}), appeared nearly thirty
years ago as Eq.~(50) of Ref.~\cite{jakeman68a}.  It was
subsequently tested experimentally in Ref.~\cite{jakeman68b}. This
supports our theory of SVS, which seems both simpler and more
general than that of Ref.~\cite{jakeman68a}. Our approach applies
for any form of $g_1(\tau)$, not just for a Lorentzian spectrum,
and it also accounts for any number of speckles per pixel.  To our
knowledge, Eqs.~(\ref{v234}a-c) have not previously appeared in
the literature.

The special case of a Lorentzian spectrum was also considered in
Ref.~\cite{BriersOC81}, which is widely cited as a founding paper
in the field of laser-speckle flowmetry. There the visibility of a
speckle pattern is quantified by ``speckle contrast'':
\begin{equation}
    K(T) \equiv \sigma_T/\langle I\rangle,
\label{K}
\end{equation}
where $\sigma_T$ is the standard deviation of the set of
intensities as measured over an exposure of duration $T$. This
quantity equals the square root of our reduced variance,
$K(T)=\sqrt{v_2(T)}$. The quoted result for a Lorentzian spectrum,
Eq.~(9) of Ref.~\cite{BriersOC81} and Eq.~(13) of a more recent
review \cite{BriersPM01}, would give $v_2(T)$ as the following
unweighted average of $[g_1(\tau)]^2$ over the exposure interval:
\begin{equation}
    \int_0^T[g_1(t)]^2{\rm d}t/T
    = {1-e^{-2x}\over 2x},
\label{briers}
\end{equation}
where $x=\Gamma T$ as before.  This conflicts with
Eq.~(\ref{v23exp}) here, and with Eq.~(50) of
Ref.~\cite{jakeman68a}, due to absence of the factors $\beta$ and
$2(1-t/T)$ in Eq.~(\ref{V2}).  The latter mistake of
Ref.~\cite{BriersOC81} is that the variance is taken as a single
integral of $[g_1(\tau)]^2$ over the exposure window $0<\tau<T$,
rather than as a double integral where $\tau=t_1-t_2$ ranges over
possible time differences within the window. The former mistake is
that the value of $\beta$, in effect, is taken as one; this is
correct only if both the pixel size is infinitesimal compared to
speckle size and if just one polarization mode is detected.  A
sampling of papers that invoke Eq.~(9) of Ref.~\cite{BriersOC81}
simultaneously match pixel size to speckle size but neglect an
unknown visibility reduction that results.  The combined error
introduced by the incorrect weighting and the neglect of $\beta$
depend on details of the experiment, but can easily exceed a
factor of ten. Hence these issues may well contribute to the
inability in the field of laser-speckle flowmetry to make
reproducible quantitative connection between speckle visibility
and blood flow speed.

\section{Demonstration of SVS}

In this section we both demonstrate the SVS technique and compare
the experimental results with theoretical predictions of the
previous section.

\subsection{Colloidal Particles}

\begin{figure} 
\includegraphics[width=\columnwidth]{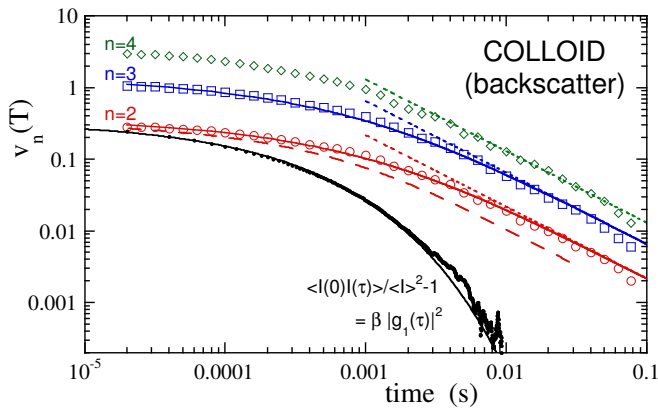}
\caption{Reduced moments of the speckle pattern vs the exposure
duration $T$, for light diffusely backscattering from an opaque
colloidal suspension.  Intensity autocorrelation data and fit from
Fig.~\protect{\ref{Ig1}} are included for comparison. The solid
curves for $n=\{2,3\}$ are generated numerically from the
$g_1(\tau)$ data in Fig.~\protect\ref{Ig1} according to our theory
of SVS, Eq.~(\protect{\ref{v234}}). The dotted lines represent the
expected long-exposure behavior, $v_n(T)\sim1/T$ of
Eq.~(\protect\ref{vnlong}).  The dashed curve is
Eq.~(\protect\ref{BriersBack}), the expectation for $v_2(T)$ based
on the formalism of Ref.~\protect\cite{BriersOC81}.}
\label{vnColloid}
\end{figure}

Our first sample is the same opaque colloidal suspension, probed
by diffusely backscattered light with the same optical setup, as
previously in Figs.~\ref{Ig1}-\ref{IT}.  For this sample, the
speckles fluctuate due to diffusion of the particles.  Now we
measure the second, third, and fourth moments of the distribution
of pixel signals, Eq.~(\ref{Sn}), and reduce the results to
dimensionless form as per Eq.~(\ref{vndef}). This is done for many
different exposure times $T$, with results shown vs $T$ in
Fig.~\ref{vnColloid}. These reduced moments appear to approach a
constant for short exposures, and to decay according to
Eq.~(\ref{vnlong}) as $1/T$ for long exposures.

To compare with expectation, we first note that $v_2(T)$ is a
weighted average of $\beta[g_1(\tau)]^2$ over the exposure
interval, Eq.~(\ref{v234}).  Therefore, we also include data for
the latter as presented previously in Fig.~\ref{Ig1}.  Recall that
the functional form for the field autocorrelation is
$g_1(\tau)=\exp(-\sqrt{\Gamma\tau})$ with
$\Gamma=6\gamma^2/\tau_o=1623~{\rm s}^{-1}$.  Evidently $v_2(T)$
and $\beta[g_1(\tau)]^2$ extrapolate to the same value at short
times, $\beta=0.34$.  But while $g_1(\tau)$ decays more rapidly,
$v_2(T)$ decays more slowly in accord with the heavy short-time
weighting in the average across the exposure interval,
Eq.~(\ref{v234}).  This qualitative agreement with expectation
also can be made quantitative. Indeed, the two solid curves in
Fig.~\ref{vnColloid} are generated by numerical integration of
$g_1(\tau)$ data according to our SVS predictions of
Eq.~(\ref{v234}).  As a check, the numerical prediction for
$v_2(T)$ matches the analytic prediction given in
Table~\ref{V2table}.  The predictions for both $v_2(T)$ and
$v_3(T)$ match the reduced moment data very well, with no
adjustable parameters.

Finally we compare with expectation based on the mistaken
formalism of Ref.~\cite{BriersOC81}.  Introducing the correct
factor of $\beta$ and taking the field autocorrelation as
$g_1(\tau)=\exp(-\sqrt{\Gamma\tau})$, the prediction for $v_2(T)$
would be
\begin{equation}
    \int_0^T \beta[g_1(t)]^2{\rm d}t/T = \beta
    {1 - (1+2\sqrt{x})\exp(-2\sqrt{x})\over 2x},
\label{BriersBack}
\end{equation}
where $x=\Gamma T$.  This is plotted as a dashed curve for
$\Gamma=1623~{\rm s}^{-1}$ and $\beta=0.34$, as known from the
intensity autocorrelation data. Evidently, the formalism of
Ref.~\cite{BriersOC81} does not correctly predict speckle
variance.

\subsection{Foam}

As another example, we collect SVS data for light diffusely
transmitted through an aqueous foam of thickness $L=1$~cm
[Gillette Foamy Regular]. The optical set-up is similar to the
colloid experiments, except that the camera is moved opposite to
the side upon which the laser light is incident.  For this
experiment, the speckles fluctuate due to sudden avalanche-like
rearrangements of bubbles within small localized sub-volumes
\cite{DJDsci91}. Such dynamics are driven by the coarsening
process, whereby small bubbles shrink and large bubbles grow in
order to lower the total interfacial surface energy
\cite{DJDpra91}.  Since the sample is far from equilibrium, and
the dynamics evolve with time, we restrict data collection to a
narrow time window centered at 100 minutes after production. Here
the average bubble diameter is $D\approx60~\mu$m, the
transport-mean free path is $l^*\approx3.5D$, and the average time
between rearrangements at each scattering site is
$\tau_o\approx20$~s \cite{DJDpra91}.  The volume of foam sampled
by the collected photons is sufficiently great that the speckle
pattern is in continuous motion.  The field correlation function
takes the same form as for DWS in transmission from a sample of
diffusing particles, $g_1(\tau) = \sqrt{6\Gamma\tau} /
\sinh\sqrt{6\Gamma\tau}$.  The first cumulant or initial decay
rate is expected to be $\Gamma=(L/l^*)^2/\tau_o \approx (1~{\rm
cm}/0.02~{\rm cm})^2/(20~{\rm s}) = 125~{\rm s}^{-1}$.  This
understanding is supported by both single-detector
\cite{DJDsci91,DJDpra91,earnshawPRE94,ADG99} and multi-speckle
\cite{HohlerPRL01,TRConFoam} dynamic light scattering experiments.

\begin{figure} 
\includegraphics[width=\columnwidth]{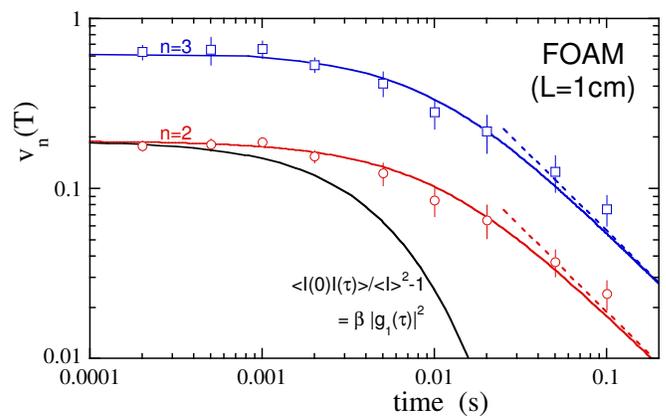}
\caption{Reduced moments of the speckle pattern vs the exposure
duration, $T$, for light diffusely transmitted through an opaque
aqueous foam of thickness $L=1$~cm.  The solid curves represent
numerical integration of the known field autocorrelation function,
$g_1(\tau) = \sqrt{6\Gamma\tau} / \sinh\sqrt{6\Gamma\tau}$,
according to the prescription of Eq.~(\protect\ref{v234}).  The
best fit is attained for $\beta=0.19$ and $\Gamma=121~{\rm
s}^{-1}$.} \label{vnFoam}
\end{figure}

Our results for the second and third reduced moments of the
distribution of pixel signals are shown in Fig.~\ref{vnFoam}.  For
short exposures both $v_2(T)$ and $v_3(T)$ approach a constant,
from which we extrapolate to zero to find $\beta=0.19$.  For
longer exposures, the moments become smaller as the speckle
pattern fluctuates more extensively during the exposure. To model
this, we numerically integrate the field correlation function
$g_1(\tau) = \sqrt{6\Gamma\tau} / \sinh\sqrt{6\Gamma\tau}$,
according to the SVS prescription of Eqs.~(\protect\ref{v234}).
Taking the first cumulant as $\Gamma=121~{\rm s}^{-1}$, close to
expectation, we obtain a satisfactory fit to both $v_2(T)$ and
$v_3(T)$ data as shown.

\section{Experimental Considerations}

This final section provides guidance on the optimal design of an
SVS experiment.  Many of the issues, and the recommendations, are
identical for other types of dynamic light scattering experiment.
Throughout we shall assume that statistics are not limited by lack
of photons. In this case it is advantageous to double the laser
power and to place a polarizer in front of the detector.  While
this does not change the average detected intensity, it does
improve the contrast in intensity levels at the plane of the
detector, and hence the signal-to-noise, since each polarization
mode forms an independent speckle pattern.  Therefore, throughout,
we shall assume that polarized detection is employed.

\subsection{Optics}

First we consider the geometry of illumination and detection.
Let $a$ be the size of the region from which emerging light is
collected.  For single-scattering experiments this could be
controlled by the diameter of the incident beam or the length it
travels within the sample.  For multiple-scattering experiments it
could be controlled by the beam diameter or the sample thickness.
The value of $a$ can also be affected by use of lenses or
apertures between the sample and the detector.  This is an
important parameter because the angular size of the speckle, in
the far field, is approximately $\lambda/a$ just as in a
diffraction experiment. Thus, if the detector is located a
distance $d$ away from the source of the collected light, then the
speckle size or spatial correlation length is approximately
$s=d\lambda/a$.

Imagining that $\lambda/a$ is fixed, and that the light intensity
can be adjusted at will, we now seek to optimize the distance $d$
at which to place the detector.  If $d$ is too small, then the
number $N_s$ of speckles at each pixel will be large and the
intercept, or maximum contrast, $\beta = v_2(0) = \langle
I^2\rangle/\langle I\rangle^2 -1$, will be small. The best case in
terms of contrast is $\beta\rightarrow1$ (or $\beta\rightarrow
1/2$ for unpolarized detection). In the opposite extreme, if $d$
is too large then each speckle will span many pixels and the
statistics of ensemble averaging will be poor. Overall, the figure
of merit to be maximized is thus $\beta\sqrt{N_s}$, the product of
maximum contrast times the spread in number of speckles per pixel.

\begin{figure} 
\includegraphics[width=\columnwidth]{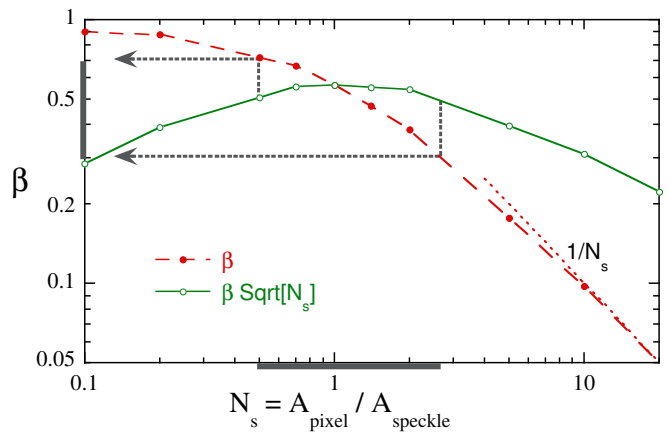}
\caption{The intercept or maximum speckle contrast, $\beta =
v_2(0) = \langle I^2\rangle / \langle I \rangle^2 -1$, and the
figure of merit $\beta\sqrt{N_s}$ to be maximized in design of
experiment, as a function of the number $N_s$ of speckles per
pixel.  An experiment with polarized detection is within about ten
percent of optimal if the intercept lies in the range
$0.3<\beta<0.7$.}\label{Beta}
\end{figure}

To find the optimal detector location by maximizing the figure of
merit requires knowing $\beta$ as a function of $N_s$.  We do this
by Monte-Carlo simulation, calculating the second intensity moment
across a specified area for speckle patterns generated at random
with the correct statistical properties. Results for $\beta$, as
well as for the figure of merit $\beta\sqrt{N_s}$, are plotted vs
$N_s$ in Fig.~\ref{Beta}.  As expected, $\beta\rightarrow 1$ for
small $N_s$ and $\beta\rightarrow 1/N_s$ for large $N_s$.  The
figure of merit achieves a maximum where the speckle size nearly
matches the pixel size, $N_s=1$.  As Fig.~\ref{Beta} demonstrates,
an experiment is within about 10\% of optimum if the intercept
lies within the range $0.3 < \beta < 0.7$ (or $0.15 < \beta <
0.35$ for unpolarized detection). Thus a good strategy is to
adjust the detector location until this criterion is met, keeping
the illumination optics fixed.  It is well-known that pixel and
speckle sizes should be matched, but to our knowledge specific
guidelines in terms of the measurable intercept have not been
published.

\subsection{Light Intensity}

Now we consider the optimal average intensity level, as controlled
by choice of laser power.  To beat photon-counting number
fluctuations and dark-count subtraction error, this power should
be as great as possible. However, high power can result in
clipping of signal for bright speckles that exceed the maximum
grayscale level of the detector. This effect introduces an error
whereby the measured intensity moments are shifted systematically
to lower values. In the opposite extreme, for low laser power, the
detected intensity levels are binned coarsely over too few
grayscale levels.  This effect introduces an error whereby the
moments are shifted systematically to higher values. Two other
effects can introduce systematic error at low laser power.  One
source is dark counts. For example, the pixels of our CCD camera
report fluctuating grayscale values of either 3 or 4, with a time
average of 3.5, when there is no illumination. The other source of
error is that the grayscale levels are reported at the lower edge
of the bin, i.e.\ $0-255$ for an 8-bit camera like ours. For
example, an actual signal level lying in the range $5\le S<6$ is
reported as a grayscale level of 5.

\begin{figure} 
\includegraphics[width=\columnwidth]{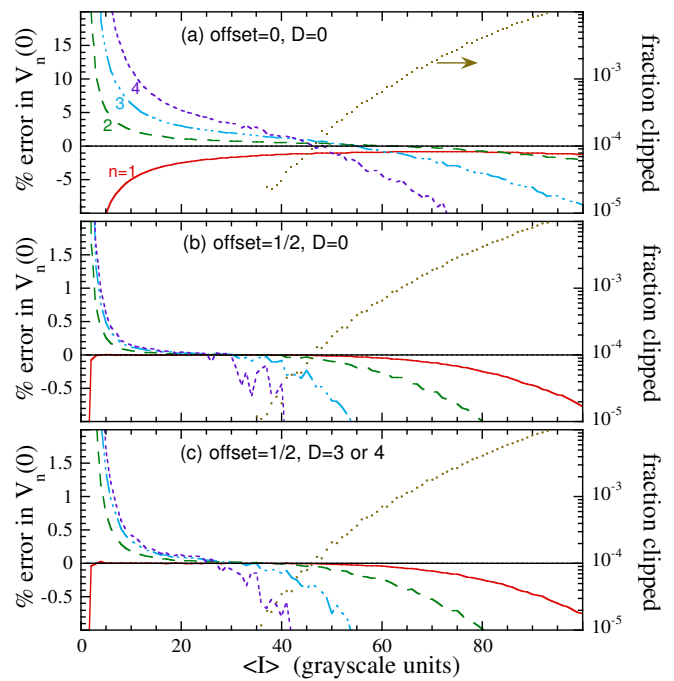}
\caption{Simulated accuracy of SVS signals, due to errors
introduced by an 8-bit digital camera, as a function of average
grayscale level. For high average intensities, the brightest
pixels are clipped to a grayscale level of 255; the fraction of
pixels that must be clipped is shown on the right-hand axis.  The
top plot shows results for moments computed directly from the
returned grayscale levels $\{0, 1, 2,\ldots, 255\}$.  The middle
shows how the accuracy is improved dramatically if pixel values
are offset by $+1/2$, which corresponds to the center of the
grayscale bin. The bottom plot differs from the middle plot by
inclusion of random dark counts, which are accounted for by
subtraction. Note that the error scales in (b-c) are ten times
smaller than the scale in (a).} \label{ErrorABC}
\end{figure}

To investigate these effects we again turn to Monte-Carlo
simulation.  At first we restrict attention to an 8-bit detector
with a pixel size of three speckles.  This gives $\beta\approx
0.3$, and hence corresponds well with our colloid experiments.  In
Fig.~\ref{ErrorABC} we display results for the systematic error in
the first four moments as a function of average intensity level.
The top plot is for an ideal detector, with zero dark counts, with
signal levels taken at the lower edge of the bins, $\{0, 1,
2,\ldots, 255\}$. Higher intensity levels are ``clipped'' to a
value of 255. The fraction of pixels that must be clipped is
plotted on the right-hand axis. At higher average intensity
levels, where clipping occurs, the intensity moments fall below
their correct values.  At lower average intensity levels, where
digitization issues occur, the intensity moments rise above their
correct values.  The middle plot shows that the latter can be
mitigated to large extent by taking the signal level at the center
of the bin. In other words, intensity moments are much more
accurate if an offset of $1/2$ is added to each reported signal,
so that possible levels are now $\{0.5, 1.5, 2.5,\ldots, 255.5\}$.
The bottom plot shows the effect of dark counts, as simulated by
randomly adding 3 or 4 to the analog signal.  This choice mimics
the conditions of our colloid experiments. To mitigate both dark
counts and lower-edge binning effect, we now subtract 3.5 from
each pixel value.  Effectively, this introduces a $\pm1$
statistical error in pixel values, which broadens the distribution
and causes higher than expected moments, as seen by comparison of
Fig.~\ref{ErrorABC}b-c. Under the operating conditions of our
colloid experiment, $\langle I\rangle=40$, we use
Fig.~\ref{ErrorABC}c to estimate that the systematic error in our
SVS data due to the combined effects of clipping, digitization,
and dark-count effects is less than 0.25\%.

\begin{figure} 
\includegraphics[width=\columnwidth]{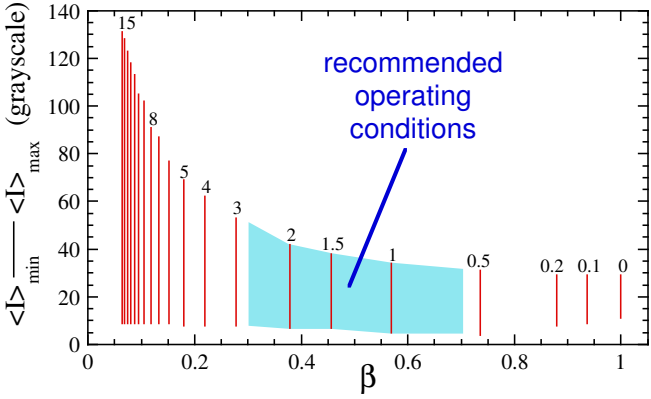}
\caption{Recommended average intensity levels for an 8-bit digital
camera, as a function of the intercept $\beta=v_2(0)$; the
corresponding pixel area is labelled in units of average speckle
area.  If the average intensity is too low, then error occurs due
to digitizing the intensity crudely into a small number of bins.
If the average intensity is too high, then error occurs due to
clipping of the high-intensity tail of the distribution.  The
vertical bars indicate the range over which these effects cause
less than a 0.1\% error in the variance.  For $n$-bit cameras, the
lower limit does not change but the upper limit scales with the
number of grayscale levels.} \label{Isafe}
\end{figure}

We now repeat the simulations for different numbers of speckles
per pixel, assuming an 8-bit camera with zero dark counts.  Plots
of error vs average grayscale level are used to identify a safe
operating range where the error in the variance is less than
0.1\%. Recommended grayscale levels are shown as a function of
intercept, $\beta$, in Fig.~\ref{Isafe}.  Once an optimal value of
$\beta$ is achieved, for example by adjusting the detector
location per the previous subsection, the laser power should be
adjusted according to this plot.  Photon-counting and dark-count
errors can be minimized by operating at the upper end of the safe
range.

\subsection{Normalization factor, $\beta$}

The perceived contrast of the intensity levels in a speckle
pattern is reduced progressively as the pixel size increases
relative to speckle size. To eliminate this effect, so that the
remaining speckle contrast can serve as a quantitative probe of
scattering site motion during the exposure window, the intercept
$\beta\equiv v_2(0)=\lim_{T\rightarrow0} (\langle I^2\rangle_T
/\langle I \rangle^2 -1)$ must be accurately determined.  One
approach, employed in our colloid and foam experiments above, is
to collect data for many exposure times and to extrapolate the
variance results to $T=0$. This is satisfactory only if the
dynamics are both stationary and sufficiently slow compared to the
fastest speed of the camera. Obviously another approach is needed
for systems with fluctuating dynamics, where each individual
exposure is to be analyzed in terms of scattering site motion at
that particular moment in time. This was the case for our first
reported application of SVS, where we probed grain motion as a
function of phase in a vibratory oscillation cycle \cite{SVS}.
There we had the luxury of being able to turn off the shaking and
to measure the contrast of the static speckle pattern under
absolutely identical illumination and detection conditions.

Here we introduce an alternative method, whereby the value of
$\beta$ can be eliminated from consideration altogether.  The idea
is to analyze not just one exposure, but rather some number $m$ of
successive exposures all of duration $T$.  The first step is to
find the variance for each of the exposures, and to average the
results together, giving $v_2(T)$. The second step is to add
together the $m$ exposures pixel-by-pixel, and to compute the
variance for the resulting ``synthetic exposure'' of duration
$mT$, giving $v_2(mT)$. These two variances depend on the value of
$\beta$, but their ratio does not:
\begin{equation}
    {v_2(mT)\over v_2(T)}
    = { \int_0^{mT} [1-t/(mT)][g_1(t)]^2
    {\rm d} t/m \over \int_0^{T} (1-t/T)[g_1(t)]^2 {\rm d} t }.
\label{vmT}
\end{equation}
The left-hand size is thus measured, and contact with scattering
site motion is made by calculation of the right-hand side for the
field autocorrelation of interest.  The predicted forms in
Table~\ref{V2table} can be used directly.  For short exposures or
slow dynamics, the variance ratios in Eq.~(\ref{vmT}) approach
one. For long exposures or fast dynamics, the variance ratios in
Eq.~(\ref{vmT}) approach $1/m$.  As a specific example, the
variance ratio for the case of a Lorentzian spectrum,
$g_1(\tau)=\exp(-\Gamma\tau)$, takes the form
\begin{eqnarray}
    {v_2(mT)\over v_2(T)}
    &=& {e^{-2mx}-1+2mx \over (e^{-2x}-1+2x)m^2 }, \\
    &\approx& { 1 +
    {12+2m+2m^2\over15(1+m)}x+{3-2m+3m^2\over15(1+m)}x^2
    \over 1 +
    {2+2m+12m^2\over15(1+m)}x+{3-2m+3m^2\over15(1+m)}mx^2
    },
\label{vmTexp}
\end{eqnarray}
where $x=\Gamma T$ as before.  The second line is a rational
approximation that is correct to $O(x^3)$ and approaches $1/m$ for
long exposures.  It can be inverted by solution of a quadratic
equation.  An additional advantage to this synthetic exposure
method is that drift in laser power or detector gain, and CCD or
CMOS noise that is correlated over successive exposures, are all
automatically cancelled.

\begin{figure} 
\includegraphics[width=\columnwidth]{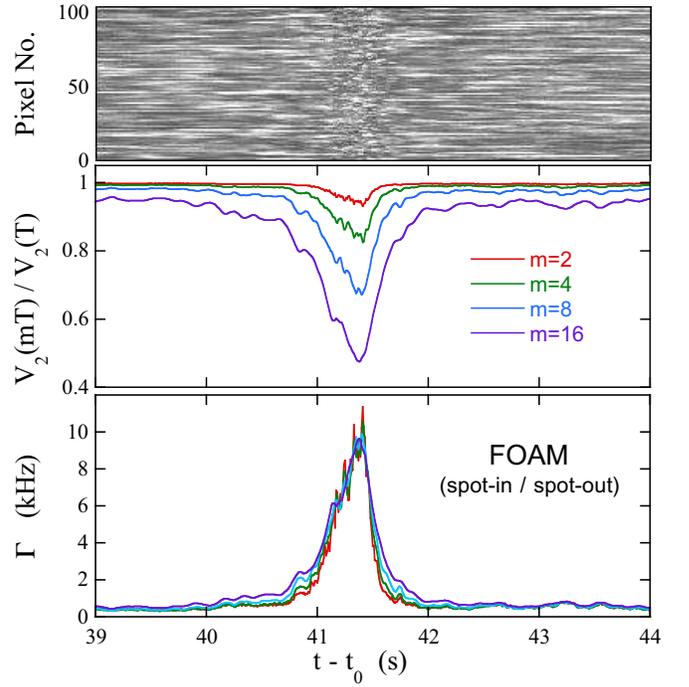}
\caption{Pixel grayscale levels, variance ratios, and Lorentzian
linewidths as a function of age for a coarsening foam. The offset
along the time axis is $t_0=6$~hrs, the lapse between sample
production and the beginning of data collection.  Here the
exposure duration is $T=10$~ms.} \label{event}
\end{figure} 

This synthetic exposure variance ratio method is now illustrated
for the same coarsening foam as in Fig.~\ref{vnFoam}.  Here the
foam is six hours old, and we employ an optical geometry whereby
photons are both introduced and collected through the same 1~mm
diameter aperture. This reduces the volume of foam sampled by the
detected photons, and ensures that only one rearrangement event is
probed at a given time.  Traditional DLS methods do not apply in
this regime.  An example rearrangement event is captured by SVS in
Fig.~\ref{event}. While the bubbles remain in a fixed location,
the speckle is nearly static and the variance ratios are nearly
one. While the bubbles move, the speckle fluctuates and the
variance ratios drop below one. According to the theory of DWS,
the spectrum is Lorentzian with linewidth $\Gamma\approx
4\pi\delta v/\lambda$, where $\delta v$ is the root-mean squared
ballistic speed of the rearranging bubbles. Analyzing the variance
ratio data using Eq.~(\ref{vmTexp}) gives nearly identical
linewidths for four different synthetic exposures, $m=\{2, 4, 8,
16\}$, as shown in the bottom plot of Fig.~\ref{event}. This good
agreement further supports our theoretical and experimental
methods of SVS. It also demonstrates how rapidly varying dynamics
may now be measured.

\begin{acknowledgments}
We thank M. Giglio, P.-A. Lemieux, R.P. Ojha, P.N. Pusey, and T.
Usher for helpful discussions.  This material is based upon work
supported by NSF under grants DMR-0305106 and PHY-0320752, and by
NASA under Microgravity Fluid Physics grant NAG3-2481.
\end{acknowledgments}

\bibliography{SVSrefs}

\end{document}